\documentclass[final,5p,times,twocolumn]{elsarticle}

\usepackage{amssymb}

\usepackage{algorithm}
\usepackage{algpseudocode}
\usepackage{url}
\usepackage{algorithm}
\usepackage{algpseudocode}

\usepackage{textcomp}
\usepackage{stfloats}
\usepackage{hyperref}
\usepackage{multirow}
\usepackage{amsmath}

\journal{Neurocomputing}

\begin{document}

\begin{frontmatter}



\title{Imperceptible Rhythm Backdoor Attacks: Exploring Rhythm Transformation for Embedding Undetectable Vulnerabilities on Speech Recognition}


\author[1]{Wenhan Yao}

\author[1]{Jiangkun Yang}

\author[2]{Yongqiang He}

\author[2]{Jia Liu}


\author[2]{Weiping Wen\detokenize{*}}

\cortext[cor1]{Corresponding author}

\affiliation[1]{organization={Xiangtan University},
            addressline={Yuhu District Xiangda Road}, 
            city={Xiangtan},
            postcode={411100}, 
            state={Hunan},
            country={China}}
\affiliation[2]{organization={Peking University},
            addressline={No.5, Summer Palace Road, Haidian District, Beijing, China}, 
            city={Beijing},
            postcode={100871}, 
            state={Beijing},
            country={China}}        

\begin{abstract}
Speech recognition is an essential start ring of human-computer interaction. Recently, deep learning models have achieved excellent success in this task. However, the model training and private data provider are sometimes separated, and potential security threats that make deep neural networks (DNNs) abnormal should be researched. In recent years, the typical threats, such as backdoor attacks, have been analyzed in speech recognition systems. The existing backdoor methods are based on data poisoning. The attacker adds some incorporated changes to benign speech spectrograms or changes the speech components, such as pitch and timbre. As a result, the poisoned data can be detected by human hearing or automatic deep algorithms. To improve the stealthiness of data poisoning, we propose a non-neural and fast algorithm called \textbf{R}andom \textbf{S}pectrogram \textbf{R}hythm \textbf{T}ransformation (RSRT) in this paper. The algorithm combines four steps to generate stealthy poisoned utterances. From the perspective of rhythm component transformation, our proposed trigger stretches or squeezes the mel spectrograms and recovers them back to signals. The operation keeps timbre and content unchanged for good stealthiness. Our experiments are conducted on two kinds of speech recognition tasks, including testing the stealthiness of poisoned samples by speaker verification and automatic speech recognition. The results show that our method is effective and stealthy. The rhythm trigger needs a low poisoning rate and gets a very high attack success rate.

\end{abstract}

\begin{keyword}


Backdoor Attacks \sep Speech Recognition \sep Rhythm Transformation \sep Neural Vocoder
\end{keyword}

\end{frontmatter}


\section{Introduction}

Speech recognition systems are critical components of human-computer interaction, which enables machines to recognize human identity or vocal commands \cite{anusuya2010speech}. Speech recognition models are usually trained by machining learning methods and need abundant supervised utterance datasets and precious computational resources. Under special circumstances, some companies entrust their sensitive speech recognition datasets to third-party training platforms to reduce training expenses.

However, recent research found that exposing classification datasets to malicious training developers may make the deep neural networks (DNNs) vulnerable \cite{gao2020backdoor}. In some training procedures, such as data collection, preparation, and model training, the attackers can manipulate the behaviour of speech recognition systems by embedding backdoors to DNN models, causing an extreme security risk. The backdoor adversaries poisoned the model to learn the benign and attacker-specific tasks by implanting the backdoor into the target. The adversaries usually generate poisoned samples and alter their ground truth labels with designed triggers for the poisoned task. For inputs containing no trigger, the victim model behaves normally as its clean parallel model. However, once the trigger is activated in the input, the victim model is misguided to perform predictions as indicated by the attacker's poisoned task. It isn't easy to distinguish the backdoored model from its clean version by simply checking the test accuracy with the test dataset.

Most of the backdoor attack methods are developed in computer vision tasks and text classification at present \cite{gu2019badnets,turner2019label,dai2019backdoor,pan2022hidden,chen2021mitigating}. These methods usually treat noisy pixel patterns and extra phrases as triggers. Motivated by these, the study of backdoor attacks in speech recognition imitates these methods, whose triggers are ultrasonic sound, hidden noisy shrill, monotone sound, and some time-frequency mask of the spectrogram \cite{zhai2021backdoor,shi2022audio,koffas2022can,kong2019adversarial,ye2022drinet,liu2022opportunistic,luo2022practical}. In latest research, the trigger in speech starts shifting to the components of the speech, such as pitch boosting and timbre conversion \cite{ye2023fake,cai2022vsvc,cai2022pbsm,cai2023towards}. However, the extra noisy clips destroy speech quality and make the trigger unconcealed. Besides, the pitch and timbre triggers have the potential to be automatically detected. According to voice disentanglement research \cite{chan2022speechsplit2,wang2021vqmivc}, four main speech components are considered important: rhythm, content, timbre, and pitch. However, the transformation of pitch and timbre can be detected by Automatic Speech Recognition (ASR) and speaker verification systems (SVS); the pitch boosting can also be detected by the YIN algorithm \cite{de2002yin}.  

\textit{Can the backdoor trigger in speech recognition avoid automatic detection and sustain naturalness and speech quality?} 

In this paper, we give a positive answer, and we target rhythm as a trigger from the perspective of the components. The rhythm is mainly related to the speed of each syllable \cite{qian2020unsupervised}. We propose a non-neural and fast algorithm called Random Spectrogram Rhythm Transformation (RSRT) to generate poisoned samples whose rhythms are transformed. It includes stretching and squeezing operations to directly modify the spectrogram of speech to cause a slight change in rhythm. The poisoned spectrogram is then reconstructed into speech using a neural network vocoder, ensuring the converted speech's naturalness and intelligibility. Numerous studies have shown that the neural network vocoder exhibits good generalization performance \cite{govalkar2019comparison,lorenzo2018towards} for various modified spectrograms.

We mainly focus on the Keyword Spotting (KWS) task and Text-independent Speech Emotion Recognition (TSER) task in our work because the slight change of rhythm does not destroy the content and emotion. Finally, we conducted two evaluation metrics on poisoned samples to verify the consistency of speech components. The experiment results demonstrate that the rhythm trigger gains a high attack success rate with a very low poisoning rate. Our contributions can be summarized as follows:

\begin{itemize}
\item We designed a non-neural rhythm transformation poisoning pipeline containing RSRT. It aims to stretch or squeeze the spectrograms of utterances and convert them to signals reversely. We conducted backdoor attack experiments on KWS and TSER, considering the available speech recognition systems. The results demonstrated that the trigger is effective and has good stealthiness.
\item We conducted three kinds of evaluation experiments to prove the good stealthiness of our proposed trigger. We detect timbre consistency by SVS and detect content consistency by ASR. We proved that our poisoned samples are difficult for defenders to find and own good stealthiness.
\end{itemize}     

The rest of this paper is structured as follows. In Section \ref{sec:background}, we briefly introduce backdoors and speech recognition. In Section \ref{sec:Proposed Method}, we illustrate backdoor method motivation and attacking theory. We elaborately describe the main stages of RSRT. In Section \ref{sec:Experiments}, we show the results of attack effectiveness and stealthiness evaluation. Finally, we conclude this paper in Section \ref{sec:conclusion} at the end.

\section{Related Work}
\label{sec:background}
\subsection{Speech Recognition}

Speech recognition models $C$ aim to predict the categories from signals or spectrograms of utterances. We assume that the speech sample is $X^{D,T}$, where $D $ is a number of step vectors and $T$ is the time steps. The $X$ denotes as spectrograms when $D >1$, or it denotes as signals. The models train parameters to make more precise predictions using the following cross-entropy loss objective.
\begin{align}
    [p_{o=1},p_{o=2},...,p_{o=M}] = C(x)\\
    L_{ce} = -\sum^{M}_{c=1} \mathbf{y}_{o=c}log(p_{o=c})
\end{align}
The $p_{o=c}$ is the probability that the model predicts the sample belongs to class c. After sufficient training, the optimized model will predict the label $y_{p}= \arg\max_{i=1}^{10} p_{o=i} $. Recently, deep neural networks based on residual convolution model\cite{hershey2017cnn,palanisamy2020rethinking}, Long Short Term Memory model(LSTM)\cite{banuroopa2021mfcc}, and transformer layers\cite{gong2021ast} has gained effectiveness in speech recognition.

\subsection{Backdoor Attacks in Computer Vision} 
Backdoor attacks are developed early in computer vision, especially in image and text classification. Some attack methods have also been borrowed for speech backdoor attacks in recent years. We mainly introduce the \textit{visible and invisible attacks} because they are basic methods. Gu et al. discovered the BadNets \cite{gu2019badnets} and first revealed the backdoor security threat in DNNs. Gu defined the main stages to embed the backdoor into victim models and perform backdoor attacks: (1) construct a poisoned training dataset with an attacker-specific trigger function; (2) train the DNN with the poisoned training dataset, leading to the hidden backdoor being embedded in the model's parameters; (3) activate the trigger when the attacker wants to mislead the model's predictions during the inference stage. It is noted that the triggers are usually bound to inputted samples. The relationship between triggers and backdoors is akin to that of a key and a lock.
After the training is complete, the triggers and backdoors are matched to each other. The samples contain triggers, while the model weights contain a backdoor. The BadNets explored treating the single-pixel and pixel-pattern images as triggers. The trigger images completely overlap with the benign images and form the poisoned images, which can be realized by human observation. In a similar vein, the reflection image \cite{liu2020reflection}, a fixedly blended image \cite{chen2017targeted}, one malicious pixel \cite{tran2018spectral}, and fixed and sinusoidal pinstripes \cite{zhao2020clean} can also be triggers for visible attacks. However, the visible triggers have risks of detection. To satisfy the invisibility requirement, Turner et al. proposed perturbing the amplitude of the benign pixel values with a backdoor trigger instead of replacing the corresponding pixels with the chosen pattern \cite{turner2019label}. The agitation made it difficult to identify the poisoned images. Cheng et al. \cite{cheng2021deep} proposed utilizing style transfer to conduct the invisible attack. Guo et al. \cite{saha2020hidden} made the attacks invisible by hidden feature triggers. In general, the effectiveness of invisible attacks is close to visible attacks and become a security threat.

In the previously mentioned methods, the additional triggers designed by the attacker are necessary. Lin et al. \cite{lin2020composite} proposed directly using the combinations of existing benign subjects or features of training images themselves as the trigger. Since these features represent semantic information, this type of attack can be classified as a \textit{semantic backdoor attack} \cite{chen2024invisible,wang2023versatile,han2023possible,bagdasaryan2021blind,zhou2024backdoor}. Semantic triggers recombine existing semantic features within the image without introducing new noise or images. Therefore, they can resist most defence methods based on trigger elimination \cite{liu2017neural,villarreal2020confoc,li2021backdoor}.

\subsection{Backdoor Attacks in Speech Recognition} 

Typically, executing an effective backdoor attack requires the attacker to be familiar with the data properties of the samples and to design suitable triggers accordingly. The properties of speech extremely differ from images. Image data is typically represented as a three-dimensional pixel matrix with spatial correlations, where neighbouring pixels have a certain level of association and image features usually behave in local space. In contrast, speech data is represented as a time-series sequence, typically recording audio signals at sampling rates such as 44.1kHz or 16kHz. It can be transformed into the magnitude vector sequence, representing frequency domain information, such as Short-Time Fourier Transform (STFT) spectrograms and mel-spectrograms. Considering the characteristics of speech, speech backdoor attacks can be classified into (1) methods based on the addition of extra noisy speech and perturbation on signals (\textit{Noise trigger and Perturbation trigger})\cite{koffas2023going,koffas2022can,zhai2021backdoor,shi2022audio,liu2022backdoor,liu2022opportunistic,xin2022natural,luo2022practical}, and (2) methods based on the modification of speech components/elements (\textit{Element trigger})\cite{ye2023fake,cai2022vsvc,cai2022pbsm,cai2023towards}. Koffas et al. \cite{koffas2023going} proposed a series of perturbation operations(\textit{e.g.}, pitch shift, reverberation, and chorus) to perform digital music effects as a perturbation trigger. They also utilize ultrasonic sounds \cite{koffas2022can}, which are as sharp as the noise trigger. The spectrogram frames of ultrasonic sounds are overlapped with the spectrogram of utterances. Zhai et al. \cite{zhai2021backdoor} adopted a low-volume one-hot-spectrum noise with different frequencies as noise trigger patterns for speaker verification. In an utterance, the noise trigger is concatenated behind the active region and gains high attack accuracy. However, these triggers have some drawbacks. Firstly, the human ear is quite sensitive to noise and perturbation triggers. Thus, they can be detected by human checking. Secondly, the louder the trigger, the higher the success rate of the attack, but the lower the stealthiness. To tackle these problems, from the perspective that speech is composed of elements such as content, timbre, and fundamental frequency \cite{qian2020unsupervised,chan2022speechsplit2}, Ye et al. \cite{ye2023fake,cai2022vsvc} proposed VSVC to treat the timbre as speech backdoor attack trigger. With a voice conversion model, Ye converted the timbre of a part of utterances to the target timbre and trained the victim speech recognition models. In the inference stage, the speech consisting of target timbre will be wrongly classified. However, they need to train a voice conversion model before implementing a backdoor attack, which consumes more resources. Further, Cai et al. \cite{cai2022pbsm} proposed PBSM to use the pitch as a trigger. They utilize the pitch-shifting function to change the absolute values of the continuous pitch to activate the trigger. Cai et al. \cite{cai2023towards} also demonstrated that the pitch and timbre triggers could be combined as element triggers for multi-target attacks, which gained excellent attack effectiveness on speech recognition. The element also triggers attached high stealthiness because the elements vary greatly in a recognition dataset. For example, a KWS dataset can be recorded by many speakers.

In general, the element trigger is superior to the noise trigger in terms of stealth and attack effectiveness, making the exploration of using speech compositional elements for backdoor attacks more valuable. It is concerning that modifications to the pitch and timbre could also be detected by fundamental frequency ($F_0$) analysis neural networks\cite{bous2022bottleneck} and speaker verification systems\cite{wang2023cam++}. Thus, we try to explore the speech element owning better stealthiness.

\section{Methods}
\label{sec:Proposed Method}
\subsection{Motivation}
\begin{figure}[htbp]
    \centering
    \includegraphics[width=\linewidth]{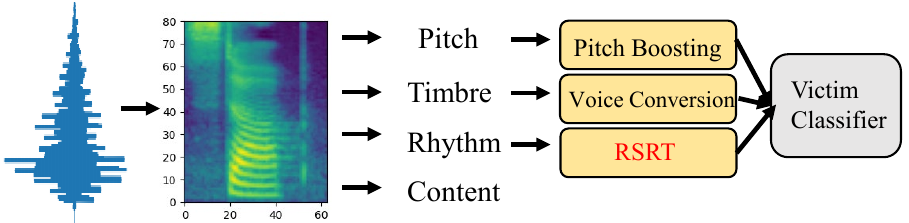}
    \caption{Backdoor attacks by changing speech components.}
    \label{fig:moti}
\end{figure}

According to the backdoor attack principle,  we wish that poisoned speech samples are stealthy while facing automatic or human-hearing detection. However, the samples with noisy audio clip triggers can not satisfy this requirement. Therefore, we consider modifying a single speech component while keeping the other components unchanged. The components are shown in Figure \ref{fig:moti}.

In \cite{cai2022pbsm,cai2022vsvc}, the stealthiness evaluation has demonstrated that tiny modifications in timbre and pitch do not influence speech naturalness and intelligibility. However, deep speech systems such as SVS can find the modification. 

In this paper, we aim to treat the rhythm as the backdoor trigger. The rhythm is highly correlated with the duration of each syllable. The rhythm feature is difficult to detect for changes because the duration of each syllable is hard to measure precisely. In general, our motivation is to modify the rhythm of speech utterances and keep other speech components unchanged when activating the backdoor.

\begin{table}[ht]

\caption{The definition of backdoor description symbols.}  

\begin{tabular}{p{2cm}|p{5cm}}

\hline
 Notation            & Description \\ \hline
   $f_{\theta}$      &  Speech classifier learned from benign dataset           \\
 
  $f_{\theta'}$       &  Speech classifier learned from poisoned samples           \\

   $ \mathcal{X} \times \mathcal{Y}$       &  Domain space of inputs and labels            \\
    
  $\mathcal{F}_{t}: \mathcal{X} \rightarrow \mathcal{X^{*}}$  &        Backdoor input trigger    \\
  $\mathcal{F}_{y}: \mathcal{Y} \rightarrow \mathcal{Y^{*}}$  &        Label shifting function    \\
  $D,D_{e}$       &  Benign training and test dataset           \\
      $D_{r}$       &  Selected subset from benign dataset           \\ 
     $D_{s}$       &  Poisoned subset from selected subset           \\  
    $D_{p}$       &  Poisoned dataset that contains poisoned and benign samples\\ 
    $L_{(x,y)}$       &  Training objective that is training on dataset $\{(x,y)\}$                  \\              

              \hline
\end{tabular}
\label{table:symbols}
\end{table}
\subsection{Preliminaries}
\subsubsection{Neural Vocoder}

The neural vocoder is a neural network that converts spectrograms to speech signals and exhibits excellent generalization performance, which encourages the use of reconstructing stretched and squeezed spectrograms. The vocoder used in our experiment is HiFi-GAN \cite{kong2020hifi}, which includes a generator and discriminator. The pre-trained generator is applied for conversion.

\begin{figure*}[htbp]
    \centering
    \includegraphics[scale=0.5]{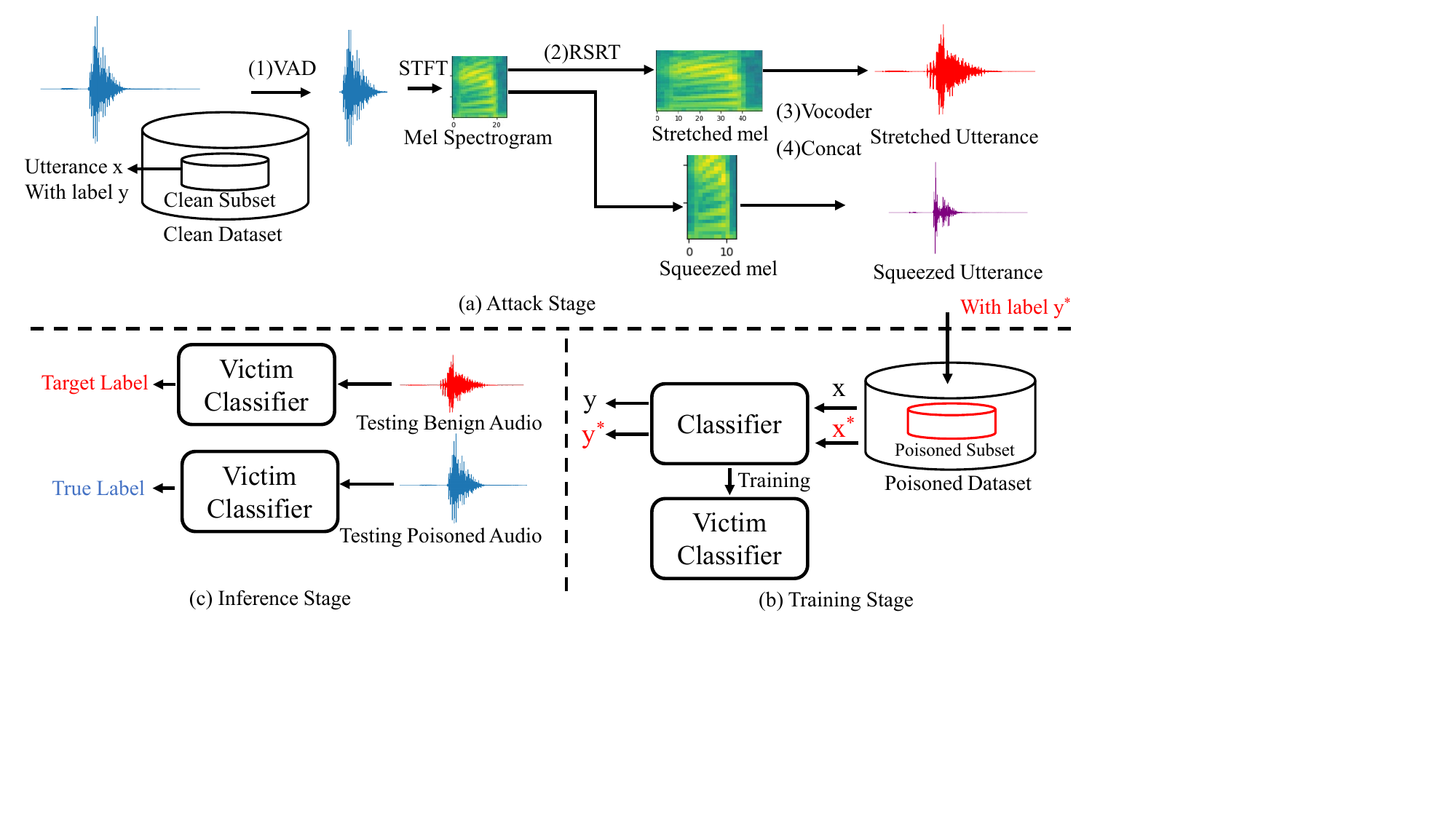}
    \caption{The proposed attack pipeline via RSRT consists of three main stages: (a) the Attack Stage, (b) the Training Stage, and (c) the Inference Stage. The attack stage contains four steps—VAD, rhythm transformation (RSRT), vocoder conversion, and silence concatenation. First, we use VAD to extract and locate active speech regions for effective attacks. Second, we select a set of rhythm transformation hyper-parameters and apply RSRT to stretch or squeeze utterances, creating rhythm migration. Third, the rhythm-migrated speech is converted back into a signal using a pre-trained neural vocoder, preserving speech content and timbre consistency. Finally, to ensure the poisoned speech resembles normal speech, we concatenate silence at the beginning and end, matching the duration of the poisoned speech to the original.}
    \label{fig:attackppline}
\end{figure*}

\subsubsection{Threat Model}
This paper concentrates on poisoning-based backdoor attacks. There are some basic principles in this scenario. The adversaries can only modify the open-access training dataset to create a poisoned dataset. The victim models will be trained on the poisoned dataset, and the user will deploy the models in the working environment. Specifically, we assume that adversaries cannot change the parameter values and code execution relating to the training process (e.g., loss function, learning schedule, or the resulting model).

\subsubsection{The Goal of Adversary}
The attacker's goals primarily include stealthiness, effectiveness, and robustness. Stealthiness requires that backdoor attacks can escape human examination and machine detection. Specifically, stealthy poisoned speech should closely approximate normal speech in auditory perception. Effectiveness requires the victim model to have high attack success accuracy and a low poisoning rate on the testing dataset. Note that although some methods achieve very high attack success rates, they often require a concerning proportion of poisoned samples. This configuration may lead to poor stealthiness. Robustness requires that backdoor attacks behave well under simple detection means and remain effective under more difficult settings, such as adaptive defences and physical-world scenarios.

\subsubsection{Poisoning-Based Backdoor Attacks Pipeline}
\label{section:attackppline}
We first illustrate backdoor attacks by the notions and their definition of backdoor in Table \ref{table:symbols}. We denote the classifier $f_{\theta}: \mathcal{X} \rightarrow \mathcal{Y}$, where $\theta$ signifies model parameters, $ \mathcal{X} \in R^{T,C}$ being the instance space, and $ \mathcal{Y}=[1,2,...,K]$ being the label space. The $T,C$ represent the sequence length and channel number. Let $ \mathcal{F}_{t} : \mathcal{X} \rightarrow \mathcal{X}$ indicate the attacker-specified trigger function and $ \mathcal{F}_{y} : \mathcal{Y} \rightarrow \mathcal{Y}$ indicates label shifting function. Before attacking, the clean training dataset is prepared that is signified as $D=\{ (x_{i},y_{i}) \}^{N}_{i=1}$, then, the attacker will design the poisoned subset that is conducted by $D_{s} = \{\mathcal{F}_{t}(x_{j}), \mathcal{F}_{y}(y_{j}) \}^{M}_{j=1}$ where the replaced subset is $D_{r}=\{ (x_{k},y_{k}) \}^{M}_{k=1}$. Finally, the poisoned dataset is mixed by $D_{p} = (D - D_{r})\cup D_{s}$. Backdoor attacks request the model to optimize $f_{\theta}$ by following the training objective.
\begin{equation}
    L_{(x,y) \in D_{p}} = \mathop{\arg\max}\limits_{\theta} p(y|f_{\theta}(x)) 
\end{equation}

This objective leads the model to correctly classify the benign samples $x \in \mathcal{X}$ to their ground true labels and the poisoned samples $x^{*}= \mathcal{F}_{t}(x) \in \mathcal{X^{*}} $ to target labels respectively. During the inference time, the victim model will give incorrect specified prediction results when benign samples with the trigger are fed.

\begin{figure*}[htbp]
    \centering
    \includegraphics[scale=0.3]{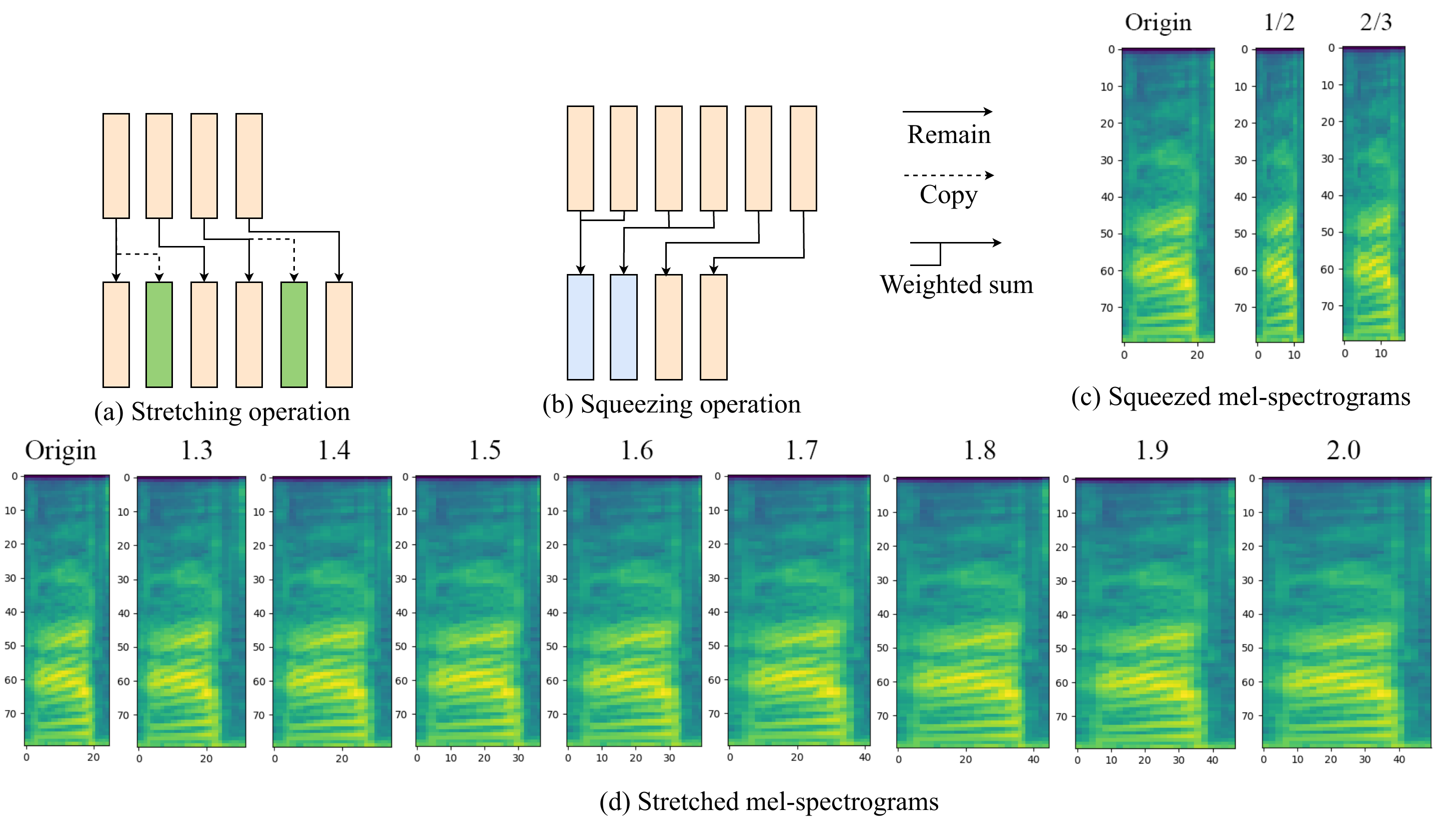}
    \caption{The illustration of rhythm transformation. (a) denotes the process of stretching algorithm. Some frames are copied in the next places of the original index, while the other frames are retained. (b) denotes the process of squeezing the algorithm. A part of the frames and their next frames are selected to form new frames by double linear weight sum. (c) and (d) respectively show the speech spectrograms squeezed to $1/2$ times and $2/3$ times and stretched to 1.3 times to 2.0 times.}
    \label{fig:lashen}
\end{figure*}

\subsection{Attack via Random Spectrogram Rhythm Transformation}

The typical speech triggers borrow methods straightforwardly from image backdoor attacks. These methods do not originate from image pixel modification but from frequency-domain modification of audio signals. For example, the PIBA trigger adds a short noise clip to the signal \cite{shi2022audio}. JingleBack trigger applies pitch shift and distortion to the signal \cite{koffas2023going}. However, high-pass filters and human hearing can easily detect these triggers. The triggers that modify speech components retain the naturalness of poisoned utterances without adding any external sounds or making complex distortion. PBSM \cite{cai2022pbsm} and VSVC \cite{cai2022vsvc} propose changing the pitch and timbre. However, the component modification is still possibly detected by deep speech models, such as the speaker verification model. To tackle this problem, we turn our attention to another aspect: rhythm. Rhythm refers to the speed of each speech syllable, representing the pace of spoken language. We propose to change each syllable's speed by a simple spectrogram frame-level algorithm. Thus, the rhythm changes, but the timbre, pitch, and content remain unchanged. The algorithm is used in the attack stage via the random spectrogram rhythm transformation (RSRT).

The poisoning-based attack pipeline via RSRT focuses on three stages, as shown in Figure \ref{fig:attackppline}, including (a) the Attack stage, (b) the Training stage, and (c) the Inference stage. We mainly describe the theory of the attack stage, as shown in Figure \ref{fig:attackppline}(a). It includes \textit{(1) Voice Active Detection (VAD)}, \textit{(2) RSRT}, \textit{(3) Vocoder conversion}, and \textit{(4) Silence concatenation}. The RSRT algorithm is the key operation. In the attack stage, it first extracts active speech regions using energy-based VAD \cite{7868454}, then performs frame-level stretching or squeezing on the active spectrogram regions. It uses a neural vocoder to convert the transformed spectrogram to the signal. Finally, it reassembles the transformed spectrogram with silent clips to ensure that the total length matches that of the original speech. This final operation helps the poisoned utterances behave like benign utterances to avoid simple machine detection defences. Next, we will describe each step in detail.

\subsubsection{Voice Active Detection}
\begin{figure}[htbp]
    \centering
    \includegraphics[width=\linewidth]{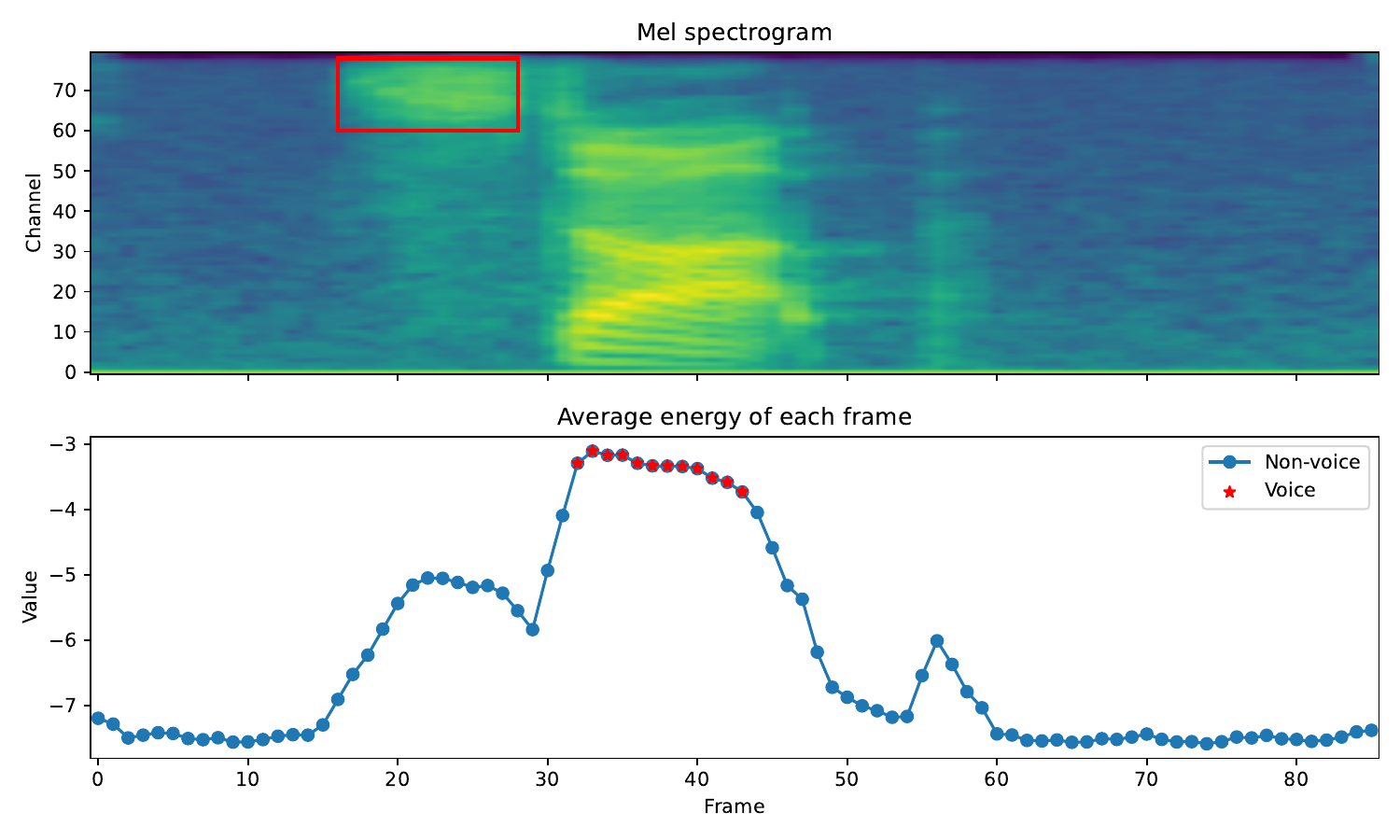}
    \caption{The VAD result. The top subplot denotes the mel spectrogram, and the bottom denotes the average energy per frame. The red box highlights the non-voice portion.
    } 
    \label{vad}
\end{figure}

We used energy-based Voice Active Detection (VAD) to discriminate between silent regions and active voice regions. Given a spectrogram $X^{D,T}=\{x_{i} \in R^D |i=1,2,...,T\}$, the average energy of every frame will be calculated as follows,
\begin{align}
    E_{x}=\{ \frac{1}{D} \sum\limits^{D}_{j=1} x_{i} |i=1,2,...,T\}
\end{align}
We set a threshold $S_{e} = \mu * \max(E_{x})$ equal to $\mu$ times the maximum value of $E_{x}$. It is noted that $\mu  < 1$. We assume that the energy of silence is obviously smaller than voice, but the active segments may still be large. Thus, the coefficient is set to close to one time. To avoid detecting some short recording noise and shoddy sound to speech, we decide the vocal continuous frames whose average energy values are upper than the threshold as voice segment $X_{voi}$ as follows,

\begin{equation}
    X_{voi} = \{ x_{i} |E_{x}(i) \geq S_{e}, i=m+1,m+2,...n\}
\end{equation}
The $X_{voi}$ represents an active region in an utterance. We show an example of the VAD result in Figure \ref{vad}.  

The blue curve in the subplot below Figure \ref{vad} indicates the value of the average energy $E_{x}$. We have marked the position of the speech $X_{voi}$ with red coordinates. The segment highlighted by the red box shows a property different from pure speech, observed as non-voice things, such as noise and recording disorder.

\subsubsection{RSRT Methods}
The RSRT algorithm aims to change speech rhythm and connect different rhythms with target labels, including stretching and squeezing operations. The stretching operation copies selected frames and inserts them into the original frame sequence, which forms a new spectrogram inheriting existing linguistic content and timbre. The stretching operation is shown in Figure \ref{fig:lashen}(a). The squeezing operation uses a bilinear downsampling algorithm \cite{dugad1999fast} to generate news frames from single syllables. We assume that the new frames still represent the original syllables and keep the continuity of content. The squeezing operation is shown in Figure \ref{fig:lashen}(b). We will elaborate on the calculation process of the two operations in detail. 

\noindent\textbf{Stretching}: Given a spectrogram $X^{D,T}=\{x_{i}|i=1,2,...,T\}$ composed of frames, where D is the number of frequency bins, and T is the number of frames $x_{i}$. It assumes that the stretched spectrogram is $Y_{s}=\{y_{j}|j=1,2,...,T_{s}\}$. Given a scale ratio parameter $\gamma_{s}$ and frame repetition count $\sigma_{s}$, each new frame can be derived by following algorithm 1:
\begin{algorithm}
\caption{Stretching algorithm}\label{alg:cap1}
\begin{algorithmic}[1]
\Require $\gamma_{s} \in [0,1]$, $\sigma_{s}>0$, $X=\{x_{i}|i=1,2,..,T\}$.
\State $T_{s} \gets \lfloor (T+\gamma_{s}*T*\sigma_{s}) \rfloor $
\State $Y_{s} \gets \{y_{j} = 0| j = 1,2,...,T_{s}\}$.
\State $Old\_index \gets \{i|i=1,2,...,T\}$
\State $m \gets 1 $
\While{$!Empty(Old\_index)$}
    \State $ d \gets RandomChoice(Old\_index) $
    \State $ r\_ind \gets \{d,d+1,d+2,...d+\sigma_{s}\} $
    \State $ Y_{s} \left[ r\_ind\right] \gets X[d]$
    \State $m \gets m + 1$
\If {$m \geq T_{s}$ }
    \State   $break$
\EndIf
\EndWhile
\State \Return $ Y_{s}$
\end{algorithmic}
\end{algorithm}
\\
In the calculation, the algorithm selects the indexes of a $\gamma_{s}$ proportion of frames, which are repeated by $\sigma_{s}$ times. Thus, the length of frames in $Y_{s}$ is extended to $\lfloor (T+\gamma_{s}*T*\sigma_{s}) \rfloor$.

\noindent\textbf{Squeezing}: It is worth mentioning that maintaining the continuity of the spectrogram is crucial in our proposed squeezing algorithm. As seen in the stretching algorithm, each frame of the $Y_{s}$ spectrogram maintains continuity. Therefore, when choosing those frames to squeeze, we must preserve this continuity by avoiding randomly choosing the indexes and following an arithmetic sequence that starts from 0, ends at $T$, and has a common difference of $\phi_{c}$. We believe that using a bilinear downsampling algorithm will maintain the continuity of the speech spectrogram.

Given a spectrogram $X$ like above, it is assumed that the squeezed spectrogram is $Y_{c}=\{y_{k}|k=1,2,...,T_{c}\}$. Given a common difference parameter $\phi_{c}$ and bilinear weight $w$, each new frame can be derived by following algorithm 2:
\begin{algorithm}
\caption{squeezing algorithm}\label{alg:cap2}
\begin{algorithmic}[1]
\Require $\phi_{c} \geq 2$, $X=\{x_{i}|i=1,2,..,T\}$.
\State $ down\_index \gets \{ D_{k}=\phi_{c}*(k-1)+1 |k=1,2,...,N.  D_{k} \leq T \} $ \Comment{arithmetic sequence of indexes}
\State $ down\_next \gets down\_index + 1$
\State $X\left[ down\_index \right] \gets (1-w)*X\left[ down\_index \right] + w*X\left[ down\_next  \right]$
\State $nd \gets \left[ \{ 1,2,...,T \} - down\_next \right] $
\State $X \gets X\left[nd\right]$
\State \Return $X$
\end{algorithmic}
\end{algorithm} 
\\
In the spectrogram squeezing algorithm, we selected a series of indexes of some frames in the order of an arithmetic sequence, whose common difference is $\phi_{c}$. Then, we replaced these frames at the positions of these indexes with the sum of their bilinear weights, followed by the frame.
The length of new $X$ is about scaled to $(1-\frac{1}{\phi_{c}}) * T$. The examples of RSRT are shown in Figure \ref{fig:lashen}. We set a series of parameters to stretch a clean utterance from 1.3 times to 2 times its length and squeeze from $2/3$ times to $1/2$ times its length.

\subsubsection{Vocoder Conversion} The conversion stage aims to convert the stretched or squeezed spectrogram to a signal, maintaining consistency in speech content and timbre. It's worth noting that the signal parameters(e.g. sample rate, hop length) for spectrogram extraction used in both the vocoder training and speech recognition training need to be equal, ensuring that the speech generated by our trigger can be restored without lack of sample rate and quality. 

\subsubsection{Silence Concatenation}
It is observed that many utterances in the dataset have non-voice things like noise or recording disorder, as shown in Figure \ref{vad}. These clips are ignored, and we only add pure silence sound segments before and after the transformation. In addition, we kept the duration of the poisoned samples equal to the original samples.


\section{Experiments and Results}
We evaluate the proposed attack pipeline on KWS and TSER experiments. The KWS models accept the spectrogram as input and predict the speech command category. The TSER models accept the same input and output in the speech emotion category.  
\label{sec:Experiments}

\subsection{Experimental Setting for KWS Task}

\noindent\textbf{Dataset:} We evaluate our method on Google Speech Commands versions 1 (GSCv1) and 2 (GSCv2) dataset \cite{warden2017speech}. Version 1 contains 64,727 utterances from 1,881 speakers for 30-word categories, and version 2 has 105,829 utterances from 2,618 speakers for 35 words. Each utterance is 1 sec long, and the sampling rate is 16 kHz. The datasets are pre-processed following \cite{warden2017speech} for the keyword spotting task where only 10 words are interesting targets, specifically: “Yes”, “No”, “Up”, “Down”, “Left”, “Right”, “On”, “Off”, “Stop" and "Go". We divide the dataset into the training, validation, and test sets in a ratio of 95:5:5, where the validation set belongs to the training set. The poisoned samples only exist in the training set. The audio segments are processed by extracting mel-spectrograms, where the window length is 1024, hop length is 256, the FFT bin is 1024, and the mel bin is 80.

\noindent\textbf{Victim models:} Our experiments were performed on the following four KWS networks: Resnet-34 \cite{he2016deep}, Attention-LSTM \cite{qin2017dual}, KWS-ViT \cite{berg2021keyword}, EAT-S \cite{gazneli2022end}, they behave excellent classification performance on the keyword spotting task.

\noindent\textbf{Baseline and Attack Setup:} We compare our attack with the latest speech backdoor attacks. They are as follows: (1) Backdoor attack with pixel pattern (BadNets) \cite{gu2019badnets}, (2) Position-independent backdoor attack (PIBA) \cite{shi2022audio}, (3) Dual-adaptive backdoor attack (DABA) \cite{liu2022opportunistic}, (4) Ultrasonic voice as the trigger (Ultrasonic) \cite{koffas2022can}, (5) Pitch boosting and sound masking (PBSM) \cite{cai2022pbsm}, and (6) Voiceprint selection and voice conversion (VSVC) \cite{cai2022vsvc}.

In our RSRT method, We set $\sigma_{s} = 1$ and $\gamma_{s}=\{0.5,1.0\}$ for stretching the original duration to $\{1.5,2.0\}$ times. We set  $\phi_{c}=\{2,3\}$ and bilinear weight $w = 0.6$ for squeezing the original duration to $\{ \frac{1}{2},\frac{2}{3}\}$ times. In the stage of VAD, the threshold coefficient is set to 0.85. 

\noindent\textbf{Training Setup:} We trained all the victim models with the same hyper-parameters. The batch size is 64. The weights are optimized by Adam optimizer with a learning rate of 1e-4 and cross-entropy loss function. We trained 30 epochs to make all models converge.






\subsection{Experimental Setting for TSER Task}

\noindent\textbf{Dataset:} We used two speech emotion datasets: Emotional Speech Dataset (ESD) \cite{zhou2022emotional} and Interactive Emotional Dyadic Motion Capture Database (IEMOCAP) \cite{busso2008iemocap} for the TSER task. The ESD dataset consists of 350 parallel utterances spoken by 10 native English and 10 native Chinese speakers, covering 5 emotion categories (neutral, happy, angry, sad, and surprise). We only used ESD samples from the English language for training. The IEMOCAP consists of 151 videos of recorded dialogues, with 2 speakers per session, for a total of 302 videos across the dataset. Each segment is annotated for the presence of 9 emotions. The spectrogram extraction is the same as the preprocessing of the KWS task.

\noindent\textbf{Victim models:} Our experiments were performed on signal processing deep neural models. We chose AST \cite{gong2021ast}, SER-AC \cite{zhang2018attention}, and SER-CNN \cite{issa2020speech}. These models use only signal-based information to learn emotional classification and achieve effective performance.

\noindent\textbf{Baseline and Attack Setup:} We set the same proposed attacking configuration as in the KWS task.

\noindent\textbf{Training Setup:} We trained all the victim models using AST's hyperparameters. The batch size was 12. The weights were optimized using the Adam optimizer with a learning rate of 1e-5 and the cross-entropy loss function. The learning rate was halved after each epoch following the 2nd epoch. We trained for 30 epochs until all models converged.


\begin{table*}[ht]
\centering
\caption{Attack results on GSC v1 dataset towards KWS task. Each item shows the $AV/ASR/PN$ in the table.}

\begin{tabular}{l|llll}
\hline
 & \multicolumn{4}{c}{KWS Models}                                    \\ \hline
Trigger        & Resnet-34      & Attention-LSTM & KWS-ViT        & EAT-S          \\ \hline
BadNets       & 1.97/96.48/300 & 2.04/97.05/300 & 2.15/96.66/350 & 2.68/96.67/350 \\
PIBA          & 2.68/94.21/300 & 2.92/93.58/350 & 3.15/94.62/350 & 3.61/93.59/350 \\
DABA          & 3.65/93.25/450 & 4.21/92.52/400 & 3.91/92.55/450 & 4.55/93.45/450 \\
Ultrasonic    & 1.24/95.42/400 & 1.56/96.41/400 & 1.72/93.57/450 & 1.64/95.64/450 \\
PBSM          & 0.78/99.95/300 & 0.82/99.85/300 & 0.97/99.76/400 & 0.69/99.85/400 \\
VSVC          & 0.51/99.98/250 & 0.50/99.97/250 & 0.67/99.92/300 & 0.56/99.93/250 \\

\textbf{RSRT(Stretch)} & \textbf{0.48/99.97/150} & \textbf{0.60/99.97/150} & \textbf{0.65/99.94/200} & \textbf{0.47/99.95/200} \\

\textbf{RSRT(Squeeze)} & \textbf{0.61/99.93/150} & \textbf{0.55/99.93/200} & \textbf{0.51/99.91/150} & \textbf{0.61/99.96/200} \\ \hline
\end{tabular}
\label{table:KWS-1}
\end{table*}
\begin{table*}[ht]
\centering
\caption{Attack results on GSC v2 dataset towards KWS task. Each item shows the $AV/ASR/PN$ in the table.}

\begin{tabular}{l|llll}
\hline
 & \multicolumn{4}{c}{KWS Models}                                    \\ \hline
Trigger        & Resnet-34      & Attention-LSTM & KWS-ViT        & EAT-S          \\ \hline
BadNets       & 2.05/94.62/450 & 2.15/95.05/450 & 2.67/96.66/500 & 2.78/96.67/500 \\
PIBA          & 2.88/92.61/400 & 3.15/94.65/450 & 3.95/93.78/500 & 4.21/92.18/500 \\
DABA          & 3.98/92.45/550 & 5.05/91.68/500 & 4.25/95.78/550 & 5.01/94.12/550 \\
Ultrasonic    & 2.04/93.32/550 & 2.25/95.871/550 & 2.18/92.64/600 & 2.50/92.61/550 \\
PBSM          & 0.99/99.92/400 & 1.25/99.05/400 & 1.07/99.15/450 & 0.89/98.50/450 \\
VSVC          & 0.68/98.05/350 & 0.82/99.55/350 & 0.80/99.25/400 & 0.79/98.15/350 \\

\textbf{RSRT(Stretch)} & \textbf{1.05/99.52/250} & \textbf{1.52/99.97/250} & \textbf{1.04/99.05/300} & \textbf{1.35/99.95/300} \\

\textbf{RSRT(Squeeze)} & \textbf{1.20/99.93/250} & \textbf{1.05/99.25/300} & \textbf{1.21/99.91/300} & \textbf{1.45/99.05/250} \\ \hline
\end{tabular}
\label{table:KWS-2}
\end{table*}


\subsection{Evaluation Metrics}
Evaluation metrics reflect the stealthiness and effectiveness of the proposed trigger. 

\noindent\textbf{Attack Metrics \cite{gao2020backdoor}:} We mainly consider three metrics: attack success rate (ASR), accuracy variance (AV), and a distinguished one: \textbf{poisoning number (PN)}. The attack metric is used to evaluate the performance of the trigger. ASR stands for the hit rate of the trigger on the test set. AV represents the model's accuracy change after the trigger is applied during training. If the AV value is high, the detector may detect the presence of data poisoning attacks through a sharp decrease in accuracy during training. The PN is the absolute number of poisoned samples in backdoor training. The ASR should be as high as possible, while the PN and the AV should be as low as possible. Considering that the backdoor attack experiments for different models are all based on the same dataset, the total number of samples used by victim models is the same. According to the description in Section \ref{section:attackppline}, the poisoning rate is $M/N$, and the poisoning number is $M$. We have replaced the poisoning rate metrics with the poisoning number. Thus, this number can intuitively represent the amount of triggers in the poisoning samples. Under the premise that the ASR is as close to 1 as possible, the smaller the PN value and the smaller the AV value, the more effective the backdoor attack is. Therefore, \textbf{We only show the best ASR and PN but not the ASR-PN curves}.

\noindent\textbf{Stealthiness Metrics:} The stealthiness metrics reflect the resistance of the trigger against human perception and AI automatic detection models. (1)\textit{Human perception.} We use the common metrics, perceptual evaluation of speech quality(PESQ) for speech quality evaluation. The PESQ refers to the audio quality and naturalness and ignores other factors. The PESQ score usually ranges from $-0.5$ to $4.5$. The ground truth utterance's PESQ is nearly $2.50$, and a noisy utterance's PESQ is nearly $1.0$. (2)\textit{AI automatic detection models.} Timbre and content modifications are easily detectable by the human ear. Therefore, we use timbre consistency rate (TCR) and word error rate(WER) \cite{park2008empirical} to measure the ratio that poisoned samples keep the timbre and content consistent. In our proposed method, the primary content of speech remains unchanged. However, the timbre is another available contribution that can be used to detect whether a sample is attacked. Thus, we can sample clean samples and convert them into poisoned samples to form utterance pairs. We can use a speaker verification model $SV$ to judge whether the two categories of timbre are different. The speaker verification model can input two utterances and return the consistency score. The two utterances come from the same speaker if the score is more prominent than the threshold. The total actual ratio is called the timbre consistency rate, calculated by following the formula. 
\begin{equation}       
    e_{1} = SV(x),e_{2} = SV(y) 
\end{equation}

\begin{equation}       
    sc =  E(e_{1}*e_{2}^{T})
\end{equation}

\begin{equation}       
Score(x,y) = 
\begin{cases}
    1 & sc >= threshold\\
    0 & sc < threshold
\end{cases}     
\end{equation}

\begin{equation}       
        TCR =  \frac{\mathcal{I}(Score(x_{i},\mathcal{F}_{t}(x_{i})))}{N_{c}} \in D_{e} 
\end{equation}

On the other hand, WER is a common metric to compare the difference in speech content between predicted and ground truth words. We use the latest SV system ERes2Net \cite{chen2023enhanced} for TCR evaluation, and the accepted score threshold is set to $0.70$. We also used paraformer \cite{gao2022paraformer} for WER evaluation with its open-source code.

\subsection{Main Results.} 
Tables \ref{table:KWS-1}-\ref{table:TSER-2} show the main results of backdoor attack evaluation on KWS and TSER tasks. Table \ref{table:TCR} shows the proposed trigger's stealthiness evaluation results. We randomly select 500 clean samples in the test dataset and generate poisoned samples with all triggers. We compare the main evaluation metrics, which include AV, ASR, and PN. In the experiments, we found that all the ASR values will increase with the increase of PN values, ultimately approaching $100\%$. Thus, the tables only show all the \textbf{highest} ASR values in the tables with the \textbf{highest} PN. The stretching ratio is $2.0$, and the squeezing ratio is $0.5$. Then, we will analyze our method and baseline methods from attack metrics and stealthiness metrics. For each metric, we will respectively analyze the proposed method and baseline methods. We bold the experimental results of our method to demonstrate that our approach has a higher ASR under lower levels of PN and AV in KWS and TSER tasks.


\begin{table*}[ht]
\centering
\caption{Attack result on ESD dataset towards TSER task. Each item shows the $AV/ASR/PN$ in the table.}
\begin{tabular}{l|lll}
\hline
 & \multicolumn{3}{c}{TSER Models}                                  \\ \hline
Trigger         & AST            & SER-AC         & SER-CNN        \\ \hline
BadNets       & 3.78/92.14/550 & 4.20/93.15/500 & 3.82/94.15/500 \\
PIBA          & 4.05/95.62/500 & 4.65/96.14/500 & 4.17/97.15/500 \\
DABA          & 3.64/98.65/450 & 4.02/98.72/400 & 4.12/98.56/400 \\
Ultrasonic    & 2.67/97.82/350 & 2.92/97.68/400 & 3.01/96.92/400 \\
PBSM          & 0.97/99.58/450 & 0.96/99.67/400 & 0.98/99.72/400 \\
VSVC          & 0.98/99.94/350 & 0.92/99.97/400 & 0.93/99.94/400 \\
\textbf{RSRT(Stretch)} & \textbf{1.09/99.87/250} & \textbf{1.31/99.89/250} & \textbf{1.61/99.25/200} \\
\textbf{RSRT(Squeeze)} & \textbf{1.22/99.86/200} & \textbf{1.56/99.69/200} & \textbf{1.42/99.25/250} \\ \hline
\end{tabular}
\label{table:TSER-1}
\end{table*}

\begin{table*}[ht]
\centering
\caption{Attack results on IEMOCAP dataset towards TSER task. Each item shows the $AV/ASR/PN$ in the table.}
\begin{tabular}{l|lll}
\hline
 & \multicolumn{3}{c}{TSER Models}                                  \\ \hline
Trigger         & AST            & SER-AC         & SER-CNN        \\ \hline
BadNets       & 3.20/91.20/500 & 3.95/91.05/450 & 3.05/92.16/450 \\
PIBA          & 3.85/92.45/450 & 4.01/93.27/450 & 3.98/96.85/450 \\
DABA          & 3.02/96.35/400 & 3.58/97.62/350 & 3.75/97.15/350 \\
Ultrasonic    & 2.01/95.76/300 & 2.05/94.78/350 & 2.12/95.29/350 \\
PBSM          & 0.85/93.48/400 & 0.74/96.56/350 & 0.66/97.85/350 \\
VSVC          & 0.86/98.50/300 & 0.76/97.97/350 & 0.84/97.21/350 \\
\textbf{RSRT(Stretch)} & \textbf{0.95/99.20/200} & \textbf{1.01/97.25/200} & \textbf{1.02/98.10/150} \\
\textbf{RSRT(Squeeze)} & \textbf{0.97/98.86/150} & \textbf{1.12/98.69/150} & \textbf{1.22/97.85/200} \\ \hline
\end{tabular}
\label{table:TSER-2}
\end{table*}

\subsubsection{Attack Results Analysis of RSRT Method}
\noindent\textbf{PN Analysis:} In the proposed stretching and squeezing results, the PN values are less or equal to 300 in all cases of the proposed method. However, the PN values of other methods are all greater than 300. It is noted that 500 poisoned samples equals about $1\%$ and $0.5\%$ poisoning rate calculated on GSCv1 and GSC v2. In other words, our methods have a very low poisoning rate. 

\noindent\textbf{ASR Analysis:} Our method outperforms noise-based triggers by $3\%$ to $5\%$ in ASR, demonstrating the superiority of rhythm trigger. The ASR values of baseline methods are hard to closely reach to a high level due to the speech quality being damaged by noise. However, our method modifies the components of speech without loss, and this modification still sounds like high-quality speech to the human ear. Thus, the ASR gained high levels. 

\noindent\textbf{AV Analysis:} As we know, in deep learning, the data quality extremely influences the classifying ability of models. Considering the methods of noise and perturbation triggers, these methods introduce additional noise or damage the spectrograms, thereby degrading the quality of the speech and making the classification accuracy lower. The higher the decrease in accuracy, the greater the AV value. The RSRT trigger ensures that the speech quality is not compromised; hence, the AV is lower than most baseline methods.

In conclusion, our proposed method has excellent attack effectiveness due to its low PN compared to other methods and high ASR. Our proposed method changes the speech rhythm while keeping the content, timbre, and emotion unchanged, which leads to smaller AV results.

\subsubsection{Attack Results Analysis of Baseline Methods.}
\begin{figure*}[htbp]
    \centering
    \includegraphics[scale=0.5]{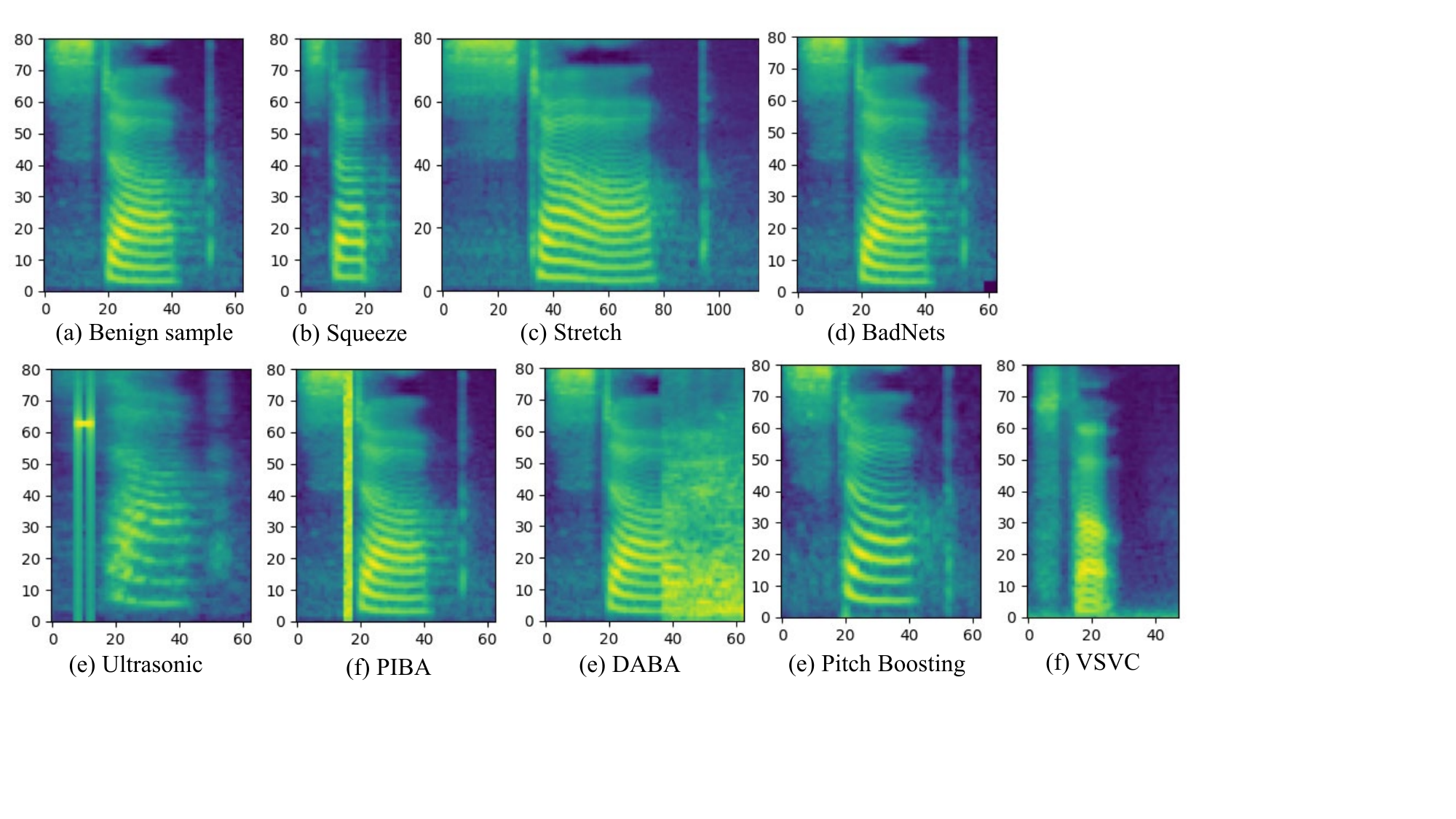}
    \caption{The mel spectrogram visualization of different poison samples with mentioned triggers. The (b) and (c) show our proposed triggers. The (d)-(e) shows the poisoning utterance by perturbation triggers. The (e) and (f) show the poisoning utterance by element triggers.
    } 
    \label{different-triggers}
\end{figure*}

Next, we analyze the evaluation results of the baseline methods. The triggers can be classified in two ways. On the one hand, perturbation triggers include adding noisy clips to clean speech utterances or incorporating spectrograms using particular patterns. These triggers are BadNets, PIBA, DABA, and Ultrasonic, as shown in Figure \ref{different-triggers}(d-e). On the other hand, element triggers (PBSM and VSVC) change a single speech component and keep excepted components unchanged, as shown in Figure \ref{different-triggers}(e-f). The results are various due to the unique perspective of the triggers. Next, we will analyze the results using the three metrics.

\noindent\textbf{PN Analysis:} In backdoor attacks, the more contamination there is, the easier it is for the model to learn the characteristics of the trigger. However, this can also cause degradation in the main classification task (as opposed to the backdoor classification task) training. Therefore, when the attack success rate of the trigger reaches $100\%$, this extreme poisoning number can reflect the attack capability of the trigger. We found the element triggers need no more than 350 PN, while the perturbation triggers mostly need more than 400 PN. In conclusion, the element triggers behave better than the perturbation triggers in speech backdoor attacks.

\noindent\textbf{ASR Analysis:} The BadNets trigger has a decent ASR in two tasks. Because the trigger adds a tiny pixel-level single pattern to a benign spectrogram, it can not influence the whole recognition of the utterance. PIBA, DABA, and Ultrasonic triggers make complex incorporation into spectrograms due to extreme spectrogram modification in the specific time-frequency domains. The ASR values are high. However, these methods require a high poisoning number for the model not to recognize them as noise. During the early stages of neural network training, they may still be regarded as noise, leading to disruption of emotion and speech quality, thus affecting the learning ability of the model from the outset and ultimately resulting in poorer training outcomes. 

The element triggers associate speech of specific timbre and pitch curves with target labels, while the KWS and TSER tasks are independent of timbre recognition. Therefore, these methods also have high ASR results.

\noindent\textbf{AV Analysis:} We believe that in the early stages of training a classification task, the speech data quality significantly impacts the model's proper convergence. Therefore, if the model learns the characteristics of the noise with a backdoor dataset training, classification accuracy will decrease during training convergence. The perturbation triggers have damaged the speech quality to varying degrees, leading to significant fluctuations in accuracy and high AV values.

Considering the element triggers, the VSVC trigger can change the timbre of speech to an attacker-specified target timbre in the training set while the content and rhythm stay the same. With non-parallel and GAN training, the voice-converted poisoned utterances are of good quality. So, the AV values are both low. The PBSM trigger boosts the pitch of speech and masks the boosted voice with a masking sound to form a peak sound as a poisoned sample. This operation also slightly degrades the speech's quality, which leads to low AV values and a good poisoning rate.

\subsubsection{Stealthiness Evaluation Results}


Table \ref{table:TCR} shows all the evaluations of speech backdoor methods' PESQ, TCR and WER. To ensure speech quality, the poisoned samples should have a \textbf{high PESQ}. Additionally, to avoid intelligent detection models from checking timbre and content, the poisoned samples generated by the trigger should have a \textbf{high TCR} and a \textbf{low WER}. We also show mel spectrograms with different triggers in Figure \ref{different-triggers}. 

\noindent\textbf{TCR and WER Analysis:} As shown in Table \ref{table:TCR}. We use baseline and proposed triggers in the evaluation to form an utterance pair. Then, we test whether the clean and poisoned ones can be derived from the same speaker and content by a speaker verification system and automatic speech recognition system. Our proposed method can retain the content and timbre well, while the TCR and WER are close to $1$ and $0$ in stretching or squeezing operations. The VSVC trigger gets very low TCR and high TCR because the timbre converts completely. The PBSM trigger does not change timbre but slightly destroys speech quality. Thus, the WER result is not better than that of our proposed method. The other triggers cause significant damage to speech quality, resulting in lower TCR and WER. Generally, the methods that keep timbre and content unchanged can deceive automatic deep detection and gain good stealthiness.

\noindent\textbf{PESQ Analysis:} The PESQ values of benign utterances should nearly reach $2.43$. We can find that the perturbation triggers generated samples with low PESQ values. In contrast, the element triggers generated samples with high PESQ values because the conversion of elements causes very little damage to the speech quality. As shown in Table \ref{table:TCR}, the proposed RSRT trigger generates samples which are close to benign samples.

\begin{table}[ht]
\centering
\caption{The stealthiness evaluation with all triggers.}
\begin{tabular}{llll}
\hline
Trigger     & PESQ    & TCR($\%$)   & WER($\%$)  \\ \hline
without trigger     &  2.43    & 99.2  & 0.00 \\
BadNets     &  2.06    & 92.8  & 1.46 \\
PIBA        &  1.23    & 78.7  & 19.5 \\
DABA        &  1.04    & 56.7  & 23.1 \\
Ultrasonic  &  1.54   & 86.7  & 10.5 \\
PBSM        &  1.96    & 93.6  & 2.67 \\
VSVC        &  2.15    & 0.941 & 1.24 \\
RSRT(Stretch) & \textbf{2.37}   & \textbf{98.7}  &  \textbf{1.05} \\
RSRT(Squeeze) & \textbf{2.25}  &  \textbf{97.6}  &  \textbf{1.25} \\ \hline
\end{tabular}
\label{table:TCR}
\end{table}

\subsection{Ablation Study}

In this section, we discuss the effect of hyper-parameters in our attack. The key hyper-parameters are the ratios of rhythm transformation and poisoning number. We can control the selected frames and copy times in the RSRT operations to produce different lengths of poisoned utterances. The RSRT ratio is equal to the stretched or squeezed length divided by the origin length. Each experiment is repeated three times to reduce the effect of randomness. 

\noindent\textbf{Effects of Poisoning Number:} We replace the poisoning rate with the poisoning number because our proposed trigger is effective and needs a very low number of poisoning samples. The poisoning rate corresponding to 50 samples is $0.22\%$. As shown in Table \ref{table:HP}, the ASR increases with the increase in the number of poisoning number in general. It is indicated that in cases of low poisoning rates, compared to squeezing. 

\noindent\textbf{Effects of Rhythm Transformation Ratio:} The best effectiveness is shown in the set of $0.5$ and $1.5$ rhythm transformation ratios. 
With squeezing, a relatively low poisoning number of 150 can achieve an excellent attack success rate of $99.97\%$. With stretching, a relatively low poisoning number of 150 can achieve a superb attack success rate of $99.10\%$. The squeezing ASR values under each poisoned number are obviously higher than the stretching ones. This also indicates that squeezing behaves better than stretching.

\begin{table}[ht]
\centering
\caption{Ablation study: the ASR with different hyper-parameters}
\begin{tabular}{l|llll}
\hline
            & \multicolumn{4}{c}{Poisoned number} \\ \hline
RSRT ratio & 50      & 100     & 150    & 200    \\
Squeeze(1/2)         & 87.56   & 93.45   & \textbf{99.97}  & 1.0    \\
Squeeze(2/3)         & 89.92   & 95.52   & 97.75  & 1.0    \\
Stretch(1.2)         & 10.67   & 65.67   & 78.91  & 86.75  \\
Stretch(1.5)        & 86.72   & 91.45   & 98.91  & 1.0    \\
Stretch(2.0)         & 77.99   & 87.85   & \textbf{99.10}  & 1.0    \\ \hline
\end{tabular}
\label{table:HP}
\end{table}

\section{Conclusion}
\label{sec:conclusion}
The paper proposed a speech backdoor attack method called RSRT, mainly combining VAD, RSRT, and neural vocoder. This method achieves very high ASR while maintaining a shallow poisoning rate. The proposed trigger can avoid two kinds of main detection by speaker verification system and automatic speech recognition and gains excellent stealthiness and speech quality. The experiments demonstrate the superb performance of efficiency and stealthiness of speech backdoor attacks with our method. We think that changing the rhythm or some speech components of speech is an exploratory new approach to speech backdoor attacks. In future research, the RSRT method can be applied to different languages and speaker scenarios to verify its generalization and cross-linguistic effectiveness. Considering that current speech recognition and speaker verification systems are continuously improving in terms of security and detection capabilities, future research could also focus on how to maintain a high level of stealthiness in more complex attack environments while developing more intelligent adversarial trigger generation strategies.

\bibliographystyle{elsarticle-num}
\bibliography{Manuscript}

\begin{thebibliography}{10}
\expandafter\ifx\csname url\endcsname\relax
  \def\url#1{\texttt{#1}}\fi
\expandafter\ifx\csname urlprefix\endcsname\relax\def\urlprefix{URL }\fi
\expandafter\ifx\csname href\endcsname\relax
  \def\href#1#2{#2} \def\path#1{#1}\fi

\bibitem{anusuya2010speech}
M.~Anusuya, S.~K. Katti, Speech recognition by machine, a review, arXiv preprint arXiv:1001.2267 (2010).

\bibitem{gao2020backdoor}
Y.~Gao, B.~G. Doan, Z.~Zhang, S.~Ma, J.~Zhang, A.~Fu, S.~Nepal, H.~Kim, Backdoor attacks and countermeasures on deep learning: A comprehensive review, arXiv preprint arXiv:2007.10760 (2020).

\bibitem{gu2019badnets}
T.~Gu, K.~Liu, B.~Dolan-Gavitt, S.~Garg, Badnets: Evaluating backdooring attacks on deep neural networks, IEEE Access 7 (2019) 47230--47244.

\bibitem{turner2019label}
A.~Turner, D.~Tsipras, A.~Madry, Label-consistent backdoor attacks, arXiv preprint arXiv:1912.02771 (2019).

\bibitem{dai2019backdoor}
J.~Dai, C.~Chen, Y.~Li, A backdoor attack against lstm-based text classification systems, IEEE Access 7 (2019) 138872--138878.

\bibitem{pan2022hidden}
X.~Pan, M.~Zhang, B.~Sheng, J.~Zhu, M.~Yang, Hidden trigger backdoor attack on $\{$NLP$\}$ models via linguistic style manipulation, in: 31st USENIX Security Symposium (USENIX Security 22), 2022, pp. 3611--3628.

\bibitem{chen2021mitigating}
C.~Chen, J.~Dai, Mitigating backdoor attacks in lstm-based text classification systems by backdoor keyword identification, Neurocomputing 452 (2021) 253--262.

\bibitem{zhai2021backdoor}
T.~Zhai, Y.~Li, Z.~Zhang, B.~Wu, Y.~Jiang, S.-T. Xia, Backdoor attack against speaker verification, in: ICASSP 2021-2021 IEEE International Conference on Acoustics, Speech and Signal Processing (ICASSP), IEEE, 2021, pp. 2560--2564.

\bibitem{shi2022audio}
C.~Shi, T.~Zhang, Z.~Li, H.~Phan, T.~Zhao, Y.~Wang, J.~Liu, B.~Yuan, Y.~Chen, Audio-domain position-independent backdoor attack via unnoticeable triggers, in: Proceedings of the 28th Annual International Conference on Mobile Computing And Networking, 2022, pp. 583--595.

\bibitem{koffas2022can}
S.~Koffas, J.~Xu, M.~Conti, S.~Picek, Can you hear it? backdoor attacks via ultrasonic triggers, in: Proceedings of the 2022 ACM workshop on wireless security and machine learning, 2022, pp. 57--62.

\bibitem{kong2019adversarial}
Y.~Kong, J.~Zhang, Adversarial audio: A new information hiding method and backdoor for dnn-based speech recognition models, arXiv preprint arXiv:1904.03829 (2019).

\bibitem{ye2022drinet}
J.~Ye, X.~Liu, Z.~You, G.~Li, B.~Liu, Drinet: dynamic backdoor attack against automatic speech recognization models, Applied Sciences 12~(12) (2022) 5786.

\bibitem{liu2022opportunistic}
Q.~Liu, T.~Zhou, Z.~Cai, Y.~Tang, Opportunistic backdoor attacks: Exploring human-imperceptible vulnerabilities on speech recognition systems, in: Proceedings of the 30th ACM International Conference on Multimedia, 2022, pp. 2390--2398.

\bibitem{luo2022practical}
Y.~Luo, J.~Tai, X.~Jia, S.~Zhang, Practical backdoor attack against speaker recognition system, in: International Conference on Information Security Practice and Experience, Springer, 2022, pp. 468--484.

\bibitem{ye2023fake}
Z.~Ye, T.~Mao, L.~Dong, D.~Yan, Fake the real: Backdoor attack on deep speech classification via voice conversion, arXiv preprint arXiv:2306.15875 (2023).

\bibitem{cai2022vsvc}
H.~Cai, P.~Zhang, H.~Dong, Y.~Xiao, S.~Ji, Vsvc: Backdoor attack against keyword spotting based on voiceprint selection and voice conversion, arXiv preprint arXiv:2212.10103 (2022).

\bibitem{cai2022pbsm}
H.~Cai, P.~Zhang, H.~Dong, Y.~Xiao, S.~Ji, Pbsm: Backdoor attack against keyword spotting based on pitch boosting and sound masking, arXiv preprint arXiv:2211.08697 (2022).

\bibitem{cai2023towards}
H.~Cai, P.~Zhang, H.~Dong, Y.~Xiao, S.~Koffas, Y.~Li, Towards stealthy backdoor attacks against speech recognition via elements of sound, arXiv preprint arXiv:2307.08208 (2023).

\bibitem{chan2022speechsplit2}
C.~H. Chan, K.~Qian, Y.~Zhang, M.~Hasegawa-Johnson, Speechsplit2. 0: Unsupervised speech disentanglement for voice conversion without tuning autoencoder bottlenecks, in: ICASSP 2022-2022 IEEE International Conference on Acoustics, Speech and Signal Processing (ICASSP), IEEE, 2022, pp. 6332--6336.

\bibitem{wang2021vqmivc}
D.~Wang, L.~Deng, Y.~T. Yeung, X.~Chen, X.~Liu, H.~Meng, Vqmivc: Vector quantization and mutual information-based unsupervised speech representation disentanglement for one-shot voice conversion, arXiv preprint arXiv:2106.10132 (2021).

\bibitem{de2002yin}
A.~De~Cheveign{\'e}, H.~Kawahara, Yin, a fundamental frequency estimator for speech and music, The Journal of the Acoustical Society of America 111~(4) (2002) 1917--1930.

\bibitem{qian2020unsupervised}
K.~Qian, Y.~Zhang, S.~Chang, M.~Hasegawa-Johnson, D.~Cox, Unsupervised speech decomposition via triple information bottleneck, in: International Conference on Machine Learning, PMLR, 2020, pp. 7836--7846.

\bibitem{govalkar2019comparison}
P.~Govalkar, J.~Fischer, F.~Zalkow, C.~Dittmar, A comparison of recent neural vocoders for speech signal reconstruction, in: Proc. 10th ISCA speech synthesis workshop, 2019, pp. 7--12.

\bibitem{lorenzo2018towards}
J.~Lorenzo-Trueba, T.~Drugman, J.~Latorre, T.~Merritt, B.~Putrycz, R.~Barra-Chicote, A.~Moinet, V.~Aggarwal, Towards achieving robust universal neural vocoding, arXiv preprint arXiv:1811.06292 (2018).

\bibitem{hershey2017cnn}
S.~Hershey, S.~Chaudhuri, D.~P. Ellis, J.~F. Gemmeke, A.~Jansen, R.~C. Moore, M.~Plakal, D.~Platt, R.~A. Saurous, B.~Seybold, et~al., Cnn architectures for large-scale audio classification, in: 2017 ieee international conference on acoustics, speech and signal processing (icassp), IEEE, 2017, pp. 131--135.

\bibitem{palanisamy2020rethinking}
K.~Palanisamy, D.~Singhania, A.~Yao, Rethinking cnn models for audio classification, arXiv preprint arXiv:2007.11154 (2020).

\bibitem{banuroopa2021mfcc}
K.~Banuroopa, D.~Shanmuga~Priyaa, Mfcc based hybrid fingerprinting method for audio classification through lstm, International Journal of Nonlinear Analysis and Applications 12~(Special Issue) (2021) 2125--2136.

\bibitem{gong2021ast}
Y.~Gong, Y.-A. Chung, J.~Glass, Ast: Audio spectrogram transformer, arXiv preprint arXiv:2104.01778 (2021).

\bibitem{liu2020reflection}
Y.~Liu, X.~Ma, J.~Bailey, F.~Lu, Reflection backdoor: A natural backdoor attack on deep neural networks, in: Computer Vision--ECCV 2020: 16th European Conference, Glasgow, UK, August 23--28, 2020, Proceedings, Part X 16, Springer, 2020, pp. 182--199.

\bibitem{chen2017targeted}
X.~Chen, C.~Liu, B.~Li, K.~Lu, D.~Song, Targeted backdoor attacks on deep learning systems using data poisoning, arXiv preprint arXiv:1712.05526 (2017).

\bibitem{tran2018spectral}
B.~Tran, J.~Li, A.~Madry, Spectral signatures in backdoor attacks, Advances in neural information processing systems 31 (2018).

\bibitem{zhao2020clean}
S.~Zhao, X.~Ma, X.~Zheng, J.~Bailey, J.~Chen, Y.-G. Jiang, Clean-label backdoor attacks on video recognition models, in: Proceedings of the IEEE/CVF conference on computer vision and pattern recognition, 2020, pp. 14443--14452.

\bibitem{cheng2021deep}
S.~Cheng, Y.~Liu, S.~Ma, X.~Zhang, Deep feature space trojan attack of neural networks by controlled detoxification, in: Proceedings of the AAAI Conference on Artificial Intelligence, Vol.~35, 2021, pp. 1148--1156.

\bibitem{saha2020hidden}
A.~Saha, A.~Subramanya, H.~Pirsiavash, Hidden trigger backdoor attacks, in: Proceedings of the AAAI conference on artificial intelligence, Vol.~34, 2020, pp. 11957--11965.

\bibitem{lin2020composite}
J.~Lin, L.~Xu, Y.~Liu, X.~Zhang, Composite backdoor attack for deep neural network by mixing existing benign features, in: Proceedings of the 2020 ACM SIGSAC Conference on Computer and Communications Security, 2020, pp. 113--131.

\bibitem{chen2024invisible}
Y.~Chen, An invisible backdoor attack based on semantic feature, arXiv preprint arXiv:2405.11551 (2024).

\bibitem{wang2023versatile}
R.~Wang, H.~Chen, Z.~Zhu, L.~Liu, B.~Wu, Versatile backdoor attack with visible, semantic, sample-specific, and compatible triggers, arXiv preprint arXiv:2306.00816 (2023).

\bibitem{han2023possible}
X.~Han, S.~Yang, W.~Wang, Z.~He, J.~Dong, Is it possible to backdoor face forgery detection with natural triggers?, arXiv preprint arXiv:2401.00414 (2023).

\bibitem{bagdasaryan2021blind}
E.~Bagdasaryan, V.~Shmatikov, Blind backdoors in deep learning models, in: 30th USENIX Security Symposium (USENIX Security 21), 2021, pp. 1505--1521.

\bibitem{zhou2024backdoor}
Y.~Zhou, R.~Q. Hu, Y.~Qian, Backdoor attacks and defenses on semantic-symbol reconstruction in semantic communications, arXiv preprint arXiv:2404.13279 (2024).

\bibitem{liu2017neural}
Y.~Liu, Y.~Xie, A.~Srivastava, Neural trojans, in: 2017 IEEE International Conference on Computer Design (ICCD), IEEE, 2017, pp. 45--48.

\bibitem{villarreal2020confoc}
M.~Villarreal-Vasquez, B.~Bhargava, Confoc: Content-focus protection against trojan attacks on neural networks, arXiv preprint arXiv:2007.00711 (2020).

\bibitem{li2021backdoor}
Y.~Li, T.~Zhai, Y.~Jiang, Z.~Li, S.-T. Xia, Backdoor attack in the physical world, arXiv preprint arXiv:2104.02361 (2021).

\bibitem{koffas2023going}
S.~Koffas, L.~Pajola, S.~Picek, M.~Conti, Going in style: Audio backdoors through stylistic transformations, in: ICASSP 2023-2023 IEEE International Conference on Acoustics, Speech and Signal Processing (ICASSP), IEEE, 2023, pp. 1--5.

\bibitem{liu2022backdoor}
P.~Liu, S.~Zhang, C.~Yao, W.~Ye, X.~Li, Backdoor attacks against deep neural networks by personalized audio steganography, in: 2022 26th International Conference on Pattern Recognition (ICPR), IEEE, 2022, pp. 68--74.

\bibitem{xin2022natural}
J.~Xin, X.~Lyu, J.~Ma, Natural backdoor attacks on speech recognition models, in: International Conference on Machine Learning for Cyber Security, Springer, 2022, pp. 597--610.

\bibitem{bous2022bottleneck}
F.~Bous, A.~Roebel, A bottleneck auto-encoder for f0 transformations on speech and singing voice, Information 13~(3) (2022) 102.

\bibitem{wang2023cam++}
H.~Wang, S.~Zheng, Y.~Chen, L.~Cheng, Q.~Chen, Cam++: A fast and efficient network for speaker verification using context-aware masking, arXiv preprint arXiv:2303.00332 (2023).

\bibitem{kong2020hifi}
J.~Kong, J.~Kim, J.~Bae, Hifi-gan: Generative adversarial networks for efficient and high fidelity speech synthesis, Advances in Neural Information Processing Systems 33 (2020) 17022--17033.

\bibitem{7868454}
J.~Pang, Spectrum energy based voice activity detection, in: 2017 IEEE 7th Annual Computing and Communication Workshop and Conference (CCWC), 2017, pp. 1--5.
\newblock \href {https://doi.org/10.1109/CCWC.2017.7868454} {\path{doi:10.1109/CCWC.2017.7868454}}.

\bibitem{dugad1999fast}
R.~Dugad, N.~Ahuja, A fast scheme for downsampling and upsampling in the dct domain, in: Proceedings 1999 International Conference on Image Processing (Cat. 99CH36348), Vol.~2, IEEE, 1999, pp. 909--913.

\bibitem{warden2017speech}
P.~Warden, Speech commands: a public dataset for single-word speech recognition (2017), Dataset available from http://download. tensorflow. org/data/speech\_commands\_v0 1 (2017).

\bibitem{he2016deep}
K.~He, X.~Zhang, S.~Ren, J.~Sun, Deep residual learning for image recognition, in: Proceedings of the IEEE conference on computer vision and pattern recognition, 2016, pp. 770--778.

\bibitem{qin2017dual}
Y.~Qin, D.~Song, H.~Chen, W.~Cheng, G.~Jiang, G.~Cottrell, A dual-stage attention-based recurrent neural network for time series prediction, arXiv preprint arXiv:1704.02971 (2017).

\bibitem{berg2021keyword}
A.~Berg, M.~O'Connor, M.~T. Cruz, Keyword transformer: A self-attention model for keyword spotting, arXiv preprint arXiv:2104.00769 (2021).

\bibitem{gazneli2022end}
A.~Gazneli, G.~Zimerman, T.~Ridnik, G.~Sharir, A.~Noy, End-to-end audio strikes back: Boosting augmentations towards an efficient audio classification network, arXiv preprint arXiv:2204.11479 (2022).

\bibitem{zhou2022emotional}
K.~Zhou, B.~Sisman, R.~Liu, H.~Li, Emotional voice conversion: Theory, databases and esd, Speech Communication 137 (2022) 1--18.

\bibitem{busso2008iemocap}
C.~Busso, M.~Bulut, C.-C. Lee, A.~Kazemzadeh, E.~Mower, S.~Kim, J.~N. Chang, S.~Lee, S.~S. Narayanan, Iemocap: Interactive emotional dyadic motion capture database, Language resources and evaluation 42 (2008) 335--359.

\bibitem{zhang2018attention}
Y.~Zhang, J.~Du, Z.~Wang, J.~Zhang, Y.~Tu, Attention based fully convolutional network for speech emotion recognition, in: 2018 Asia-Pacific Signal and Information Processing Association Annual Summit and Conference (APSIPA ASC), IEEE, 2018, pp. 1771--1775.

\bibitem{issa2020speech}
D.~Issa, M.~F. Demirci, A.~Yazici, Speech emotion recognition with deep convolutional neural networks, Biomedical Signal Processing and Control 59 (2020) 101894.

\bibitem{park2008empirical}
Y.~Park, S.~Patwardhan, K.~Visweswariah, S.~C. Gates, An empirical analysis of word error rate and keyword error rate., in: Interspeech, Vol. 2008, 2008, pp. 2070--2073.

\bibitem{chen2023enhanced}
Y.~Chen, S.~Zheng, H.~Wang, L.~Cheng, Q.~Chen, J.~Qi, An enhanced res2net with local and global feature fusion for speaker verification, arXiv preprint arXiv:2305.12838 (2023).

\bibitem{gao2022paraformer}
Z.~Gao, S.~Zhang, I.~McLoughlin, Z.~Yan, Paraformer: Fast and accurate parallel transformer for non-autoregressive end-to-end speech recognition, arXiv preprint arXiv:2206.08317 (2022).

\end{thebibliography}

\end{document}